\documentclass[journal=jpclcd,manuscript=letter]{achemso}

\usepackage{graphicx}
\usepackage{amsmath}
\usepackage{amssymb}

\author{Eliza M. McIntosh}
\email{emb56@cam.ac.uk}
\affiliation{The Cavendish Laboratory, J.J. Thomson Avenue, Cambridge, CB3 0HE, United Kingdom}
\altaffiliation{These authors contributed equally to the work.}
\author{K. Thor Wikfeldt}
\email{wikfeldt@hi.is}
\affiliation{Thomas Young Centre, London Centre for Nanotechnology and Department of Chemistry, University College London, London WC1E 6BT, United Kingdom}
\altaffiliation{Science Institute and Faculty of Science, VR-III, University of Iceland, 107 Reykjavik, Iceland}
\altaffiliation{These authors contributed equally to the work.}
\author{John Ellis}
\affiliation{The Cavendish Laboratory, J.J. Thomson Avenue, Cambridge, CB3 0HE, United Kingdom}
\author{Angelos Michaelides}
\affiliation{Thomas Young Centre, London Centre for Nanotechnology and Department of Chemistry, University College London, London WC1E 6BT, United Kingdom}
\author{William Allison}
\affiliation{The Cavendish Laboratory, J.J. Thomson Avenue, Cambridge, CB3 0HE, United Kingdom}

\title[\texttt{achemso}]
{Quantum effects in the diffusion of hydrogen on Ru(0001)}

\begin{document}
\newpage
\begin{abstract}
An understanding of hydrogen diffusion on metal surfaces is important, not just for its role in heterogeneous catalysis and hydrogen fuel cell technology, but also because it provides model systems where tunneling can be studied under well-defined conditions. 
Here we report helium spin-echo measurements of the atomic-scale motion of hydrogen on the Ru(0001) surface between 75 and 250~K.  
Quantum effects are evident at temperatures as high as 200~K, while below 120~K we observe a tunneling-dominated temperature independent jump rate of $1.9 \times 10^{9}$~s$^{-1}$, many orders of magnitude faster than previously seen. 
Quantum transition state theory calculations based on ab initio path-integral simulations reproduce the temperature dependence of the rate at higher temperatures and predict a crossover to tunneling-dominated diffusion at low temperatures, although the tunneling rate is under-estimated, highlighting the need for future experimental and theoretical studies of hydrogen diffusion on well-defined surfaces.
\end{abstract}

\begin{figure}
  \includegraphics[width=5.1cm]{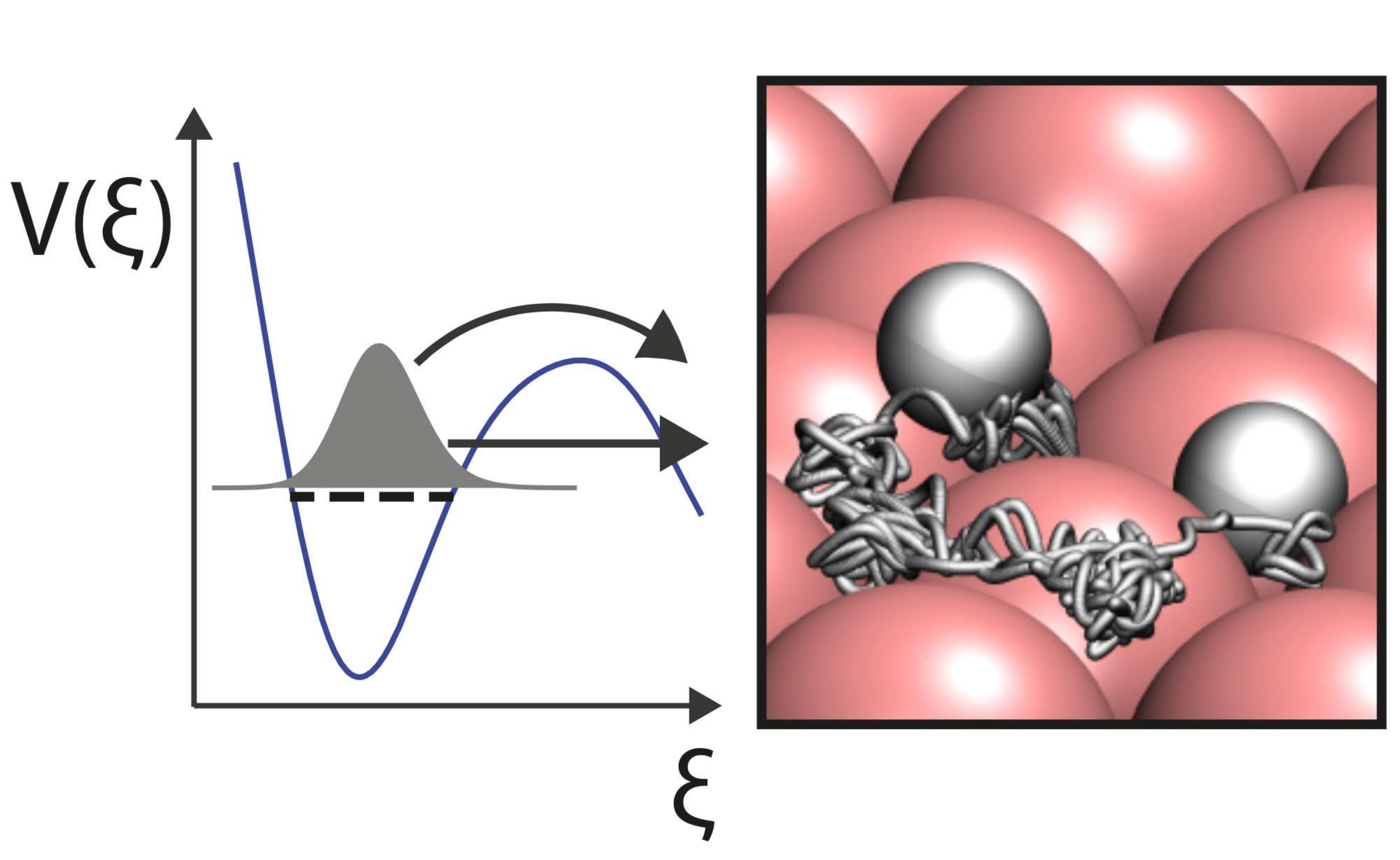}
\end{figure}


\newpage


Hydrogen (H) atoms can exhibit significant quantum nuclear effects in many and varied materials, such as H-bonded crystals and ferroelectrics~\cite{li2011quantum,garcia2004enzymes,koval2002ferro}, high pressure ice~\cite{benoit1998tunnelling}, and at the surface and in the bulk of metals \cite{li2010quantum,fukai2005metal,alanissila2002surfdiffus}, 
where the rate of diffusion is enhanced by quantum tunneling.  
Studies of H diffusion on metal surfaces provide model systems where tunneling can be studied in great detail under well-defined conditions.

Experimental measurements of H diffusion on metal surfaces are, however, extremely challenging and have at times been contentious. This is illustrated \textit{e.g.} by the case of H on Ni(111) where initially a sharp classical to quantum crossover was reported along with similar rates for H and D~\cite{gomer1991fem,lee1993observation}, whereas later experiments found only a weak crossover and normal isotope effect~\cite{cao1997HNi111}. Subsequent calculations from Badescu \textit{et al.} explained the experimental observations through a mechanism involving tunneling from excited vibrational states~\cite{badescu2001quantum}. For H on Cu(001) scanning tunneling microscopy (STM) experiments~\cite{ho2000stm} revealed a sharp crossover around 60~K to a nearly temperature independent tunneling regime, and subsequent density functional theory (DFT) calculations~\cite{kua2001direct,sundell2004HDCu001,sundell2005HDCu001} of the tunneling rate successfully reproduced the experimental results. Despite these 
studies, a detailed view of the underlying mechanisms behind the quantum behavior of H and other light adsorbates on surfaces is still lacking and well-defined comparisons between experiment and theory are uncommon. 

 
We present here a combined experimental and theoretical study of the temperature dependence and mechanism of diffusion of H on Ru(0001), a system well suited to such an approach:  widespread experimental interest exists~\cite{feulnerandmenzel,mak1986LITD,lindroos1987study,Braun1992,sandhoff1993MC,Shi1994vibrations,kostov2004hreels,tatarkhanov2008hydrogen,lizzit2009Ru0001core}, and previous DFT studies~\cite{chelikowsky1987HRu,xu2005HRu0001castep,lizzit2009Ru0001core,KristinsdottirDFT} found good agreement with experimental adsorption geometries and vibrational frequencies, suggesting that DFT provides an accurate description of the potential energy surface -- a prerequisite for reliable rate calculations.  

Experimentally, we use the novel helium Spin-Echo technique~\cite{jardine2009helium} which allows determination of surface dynamics on pico- to nano-second timescales through measurement of the Intermediate Scattering Function (ISF) $I\left(\Delta{\mathbf{K}},t\right)$, a measure of surface correlation after a time $t$ on the direction and length-scale given by $\Delta{\mathbf{K}}$.  
Our calculations combine a DFT description of the electronic structure with path-integral molecular dynamics (PIMD) to obtain
a unified description of jump rates at high and low temperatures with full account of the coupling to surface phonons. The PIMD technique provides a quantum mechanical description of configuration 
space by representing atomic nuclei as ring-polymers of interconnected beads,
and the limit of exact quantum mechanical properties can be approached by 
increasing the number of beads. To obtain jump rates including tunneling contributions 
we then combine the PIMD simulations with quantum transition state theory 
(QTST)~\cite{gillan1987quantum,voth1989qtst}.
Despite a significant computational cost, our calculations demonstrate for the first time the feasibility of investigating tunneling-assisted surface diffusion entirely from first principles using PIMD-based QTST.

We show that the diffusion of H on Ru(0001) exhibits uniquely interesting properties.  At the highest temperatures studied overbarrier hopping between non-degenerate fcc and hcp sites is seen, while a constant diffusion rate of $\sim 3 \times 10^{9}$~s$^{-1}$ below $\sim 120$~K indicates quantum tunneling at rates many orders of magnitude faster than observed for other systems: 4 decades faster than the highest temperature independent rate for H on Ni(111)~\cite{lee1993observation}, 5 decades faster than H on Cu(111)~\cite{sykes2012}, and 12 decades faster than for H on Cu(001)~\cite{ho2000stm}.
Additionally, for H on Ru(0001), the presence of multiple jumps indicates low adsorbate-substrate friction compared to \textit{e.g.} the diffusion of H on Pt(111), where HeSE was used to study quantum contributions to the activated motion~\cite{jardine2010HPt111_HeSE}.  We also find evidence for repulsive inter-adsorbate interactions.  
The overall temperature dependence of ab initio PIMD rates agrees reasonably with experiment, with good agreement down to around 200~K. At low temperatures the rate is underestimated, and we use a harmonic quantum transition state theory (HQTST) version of instanton theory~\cite{andersson2009comparison} to shed light on this discrepancy. This method is also based on Feynman's path-integral formalism 
and, in the present implementation, provides the effective lowering of the energy barrier on 
a rigid potential energy surface (PES) due to quantum tunneling; 
it thus does not treat all degrees of freedom explicitly as in PIMD simulations.

\begin{figure}
 \includegraphics[width=8.25cm]{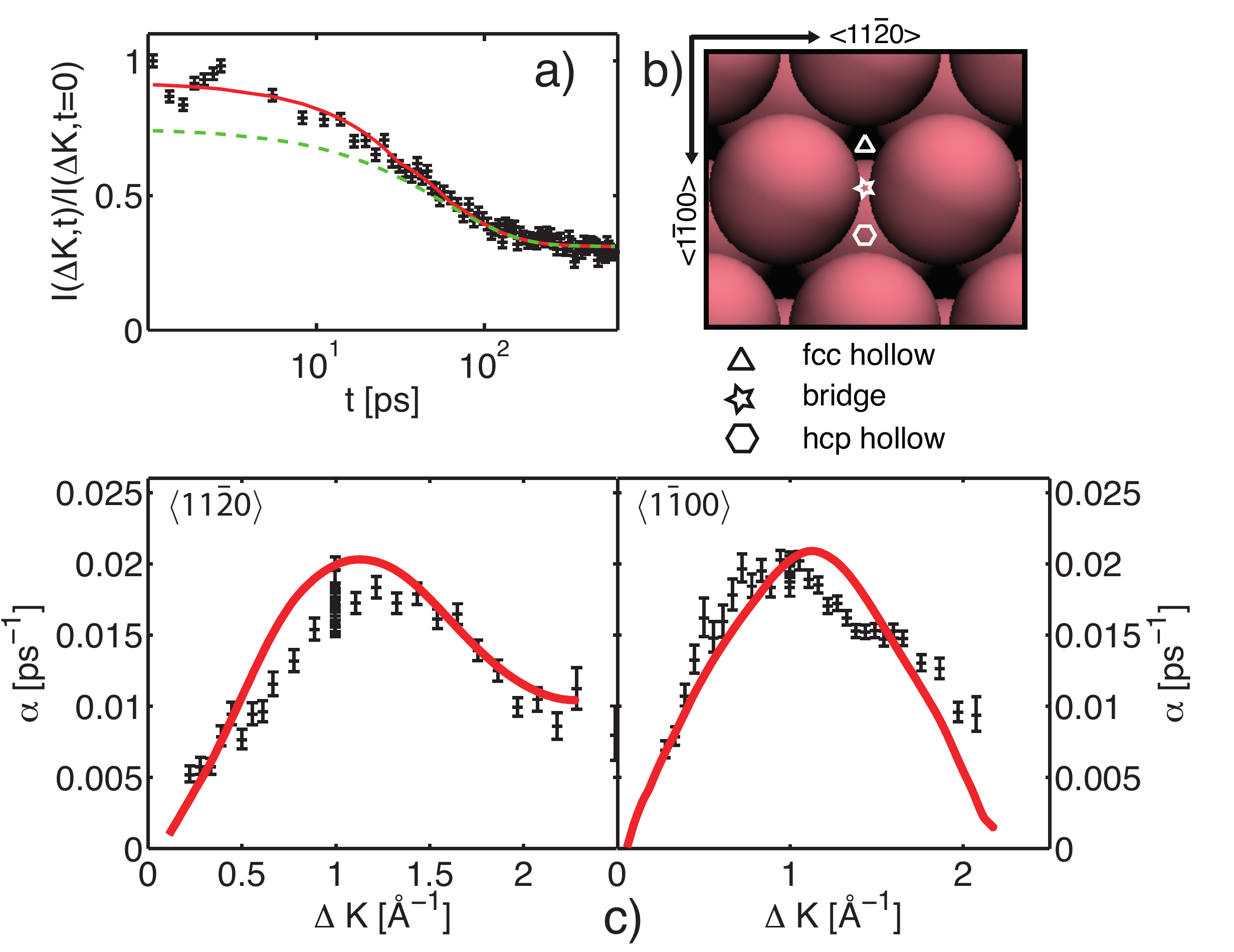}
 \caption{\label{fig:ISF} a) Example of an ISF from experiment for T~$= 250$~K, $\Delta \mathbf{K} = 1.76$~\AA$^{-1}$ along $\left\langle11\bar{2}0\right\rangle$, with the results of the best-fit Monte Carlo simulation (solid red line) and a one exponential fit to the long-time limit data (dashed green line) also shown.  b)  Schematic of the Ru(0001) surface showing the $\left<11\bar{2}0\right>$ and $\left<1\bar{1}00\right>$ directions. c)  $\alpha$--$\Delta \mathbf{K}$ plots along $\left<11\bar{2}0\right>$ (left) and $\left<1\bar{1}00\right>$ (right) at 250~K.  Data points are shown with black crosses, and the result of the best fit Monte Carlo simulations (see text) is shown with a thick red line.}
\end{figure}

\ref{fig:ISF}a) shows a typical measurement (an ISF) at 250~K, in which a loss in surface correlation is seen over time as $I\left(\Delta \mathbf{K},t\right)$ decreases.  This is attributed to diffusion of H on the Ru(0001) surface; measurements on a clean Ru(0001) surface show no such decay.  By treating each ISF as a combination of 
exponential decays $\exp(-\alpha t)$, dephasing rates ($\alpha$) can be obtained.  Two exponential decays were needed to model the experimental data at 250~K along $\left<1\bar{1}00\right>$ and $\left<11\bar{2}0\right>$;  this is indicative of hopping between two non-degenerate sites on a non-Bravais lattice \cite{tuddenham2010lineshapes}, here the fcc and hcp hollow sites.

A Bayesian method was used to fit the entire HeSE data set at 250~K (85 ISFs) with two exponential decays using the analytic model of Tuddenham et al. \cite{tuddenham2010lineshapes}, with the basic jump rate and energy difference between the fcc and hcp adsorption sites as variable parameters.  This gave a peak in the relative probability density at the most probable value of the fcc-hcp site energy difference, 22.2~meV, with a statistical uncertainty of $\pm$~0.6~meV.  \ref{fig:ISF}c) shows the variation of the slow decay constant $\alpha$ with $\Delta\mathbf{K}$.  The steep approach to the origin at low $\Delta \mathbf{K}$ is indicative of multiple jumps (low adsorbate-substrate friction) \cite{jardine2009helium}.  A dip in $\alpha$ at $\Delta\mathbf{K}$~=~$\sim$1.3~\AA$^{-1}$ is also seen (more obvious along $\left<1\bar{1}00\right>$ due to the shape of the curves), arising from repulsive inter-adsorbate interactions which stabilize ordering when 
the adsorbates are as far apart as possible.  Correlations at values of $\Delta{\mathbf{K}}$ corresponding to this separation tend to decay more slowly, giving the dip in the dephasing rate, and a corresponding diffuse ring in helium diffraction scans~\cite{supmat,jardineCs,GilNaCu}.

A Monte Carlo simulation~\cite{supmat} of hopping between fcc and hcp sites for 0.2~ML~H coverage provided quantitative insights into the diffusion mechanism and gave a second estimate of the fcc-hcp site energy difference as 18.7~meV, with a statistical uncertainty of $\pm$~0.3~meV.  Systematic uncertainty from approximations inherent in this and the analytic model~\cite{tuddenham2010lineshapes} exceed the statistical uncertainties, but we can conclude that these independent measures of the site energy difference are consistent with a value of $(20\pm5)$~meV.  The variation of $\alpha$ from the Monte Carlo ISFs with $\Delta{\mathbf{K}}$ is shown with the thick line in \ref{fig:ISF}c), which generally fits the data well.  The biggest deviation from experiment is at $\Delta\mathbf{K}\sim$1.3~\AA$^{-1}$, the position of the de-Gennes feature \cite{jardineCs}, suggesting that the form of the repulsive inter-adsorbate interaction is more complex than the simple dipole-dipole repulsion used here.

\begin{figure}
 \includegraphics[width=8.25cm]{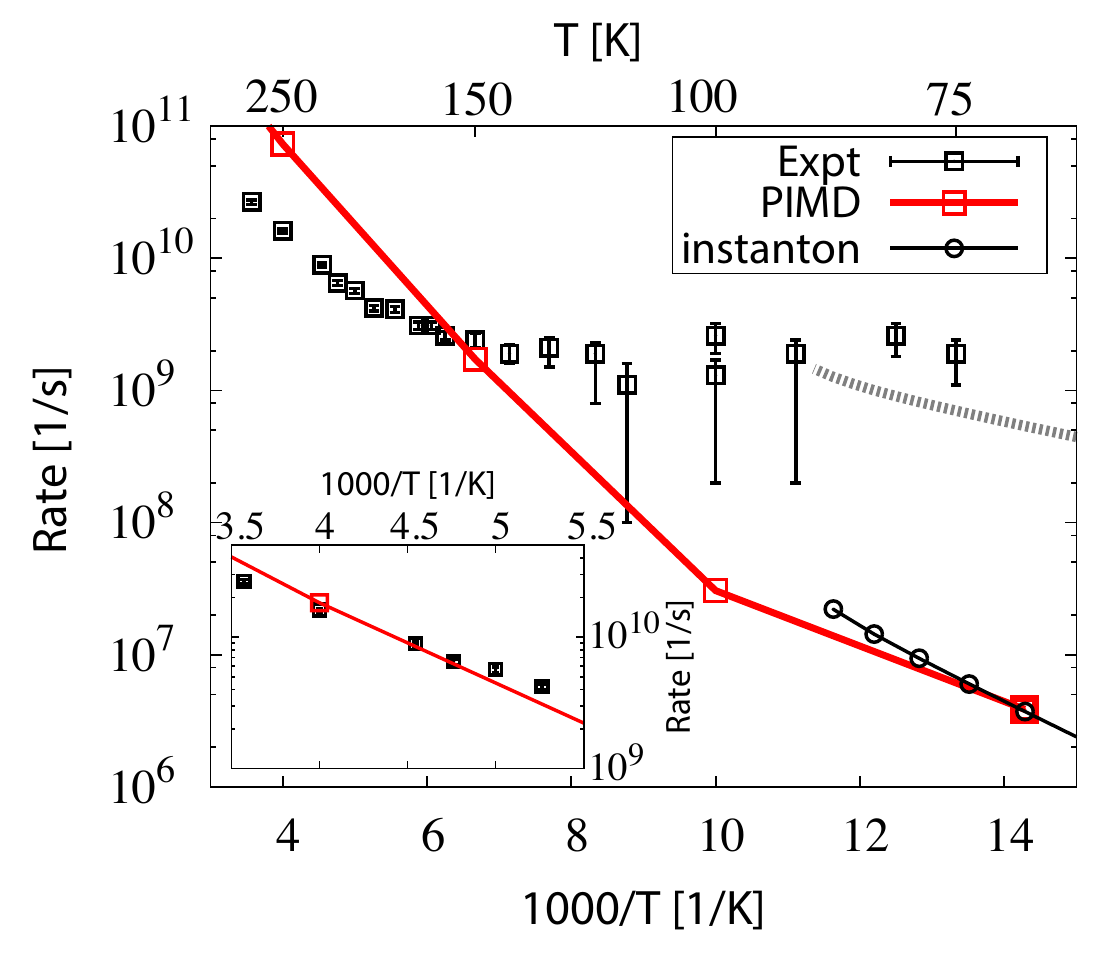}
 \caption{\label{fig:rate} Experimental rates (black squares), calculated jump rates from PIMD (red line with squares, error bars denoted by thick shaded lines) and instanton calculations on a one-dimensional potential (black line with circles). 
 The dashed line shows instanton results where the fcc-hcp energy difference is reduced to 25~meV. The inset shows high temperature PIMD rates scaled by 0.25 where PIMD results agree with the slope of the experimental data.}
 \end{figure}

\ref{fig:rate} reports the H diffusion rates measured across a broad range of temperatures, from 250~K down to 75~K. At the highest temperatures a linear dependence of the logarithm of the diffusion rate with 1/T is observed.  As the temperature is reduced the diffusion rate levels off, until below $\sim$120~K the diffusion rate appears to be independent of T, to within the experimental errors.  
Note that even the lowest rates reported in this Figure are comfortably within the experimental range of the technique, as discussed in~\cite{supmat}.
The linear dependence at high T indicates activated H diffusion, and a fit to all data shown of the form $A T \exp\left(-E_{a}/k_{B}T\right)+C$ yields an estimate of the classical activation energy for fcc-hcp hopping $E_{a} = (95\pm3)$~meV.  The pre-exponential factor is $(4.5\pm0.6)\times10^{9}$~s$^{-1}$K$^{-1}$, and the temperature-independent tunneling rate $C$ is $(1.9\pm0.1)\times10^{9}$~s$^{-1}$. 

Having obtained experimental rates for H diffusion across a broad range of temperatures we now explore the system with DFT. 
Our calculations show that the fcc site is the most stable adsorption site, with the hcp site less stable by about 50~meV which is about 30~meV larger than the experimental estimate.  
The lowest energy diffusion pathway is across the bridge site with an activation barrier of 150 meV, which reduces to 120~meV when zero-point energy (ZPE) effects are taken into account within the harmonic approximation. 
The experimentally observed rate cannot be described by classical transition state theory as quantum effects are clearly seen.
We therefore turn to ab initio PIMD, an approach that can capture the change of the quantum free energy barrier for diffusion due to the quantum nature of the proton, and thus account for tunneling and ZPE effects beyond the harmonic approximation.

\begin{figure}
 \includegraphics[width=8.25cm]{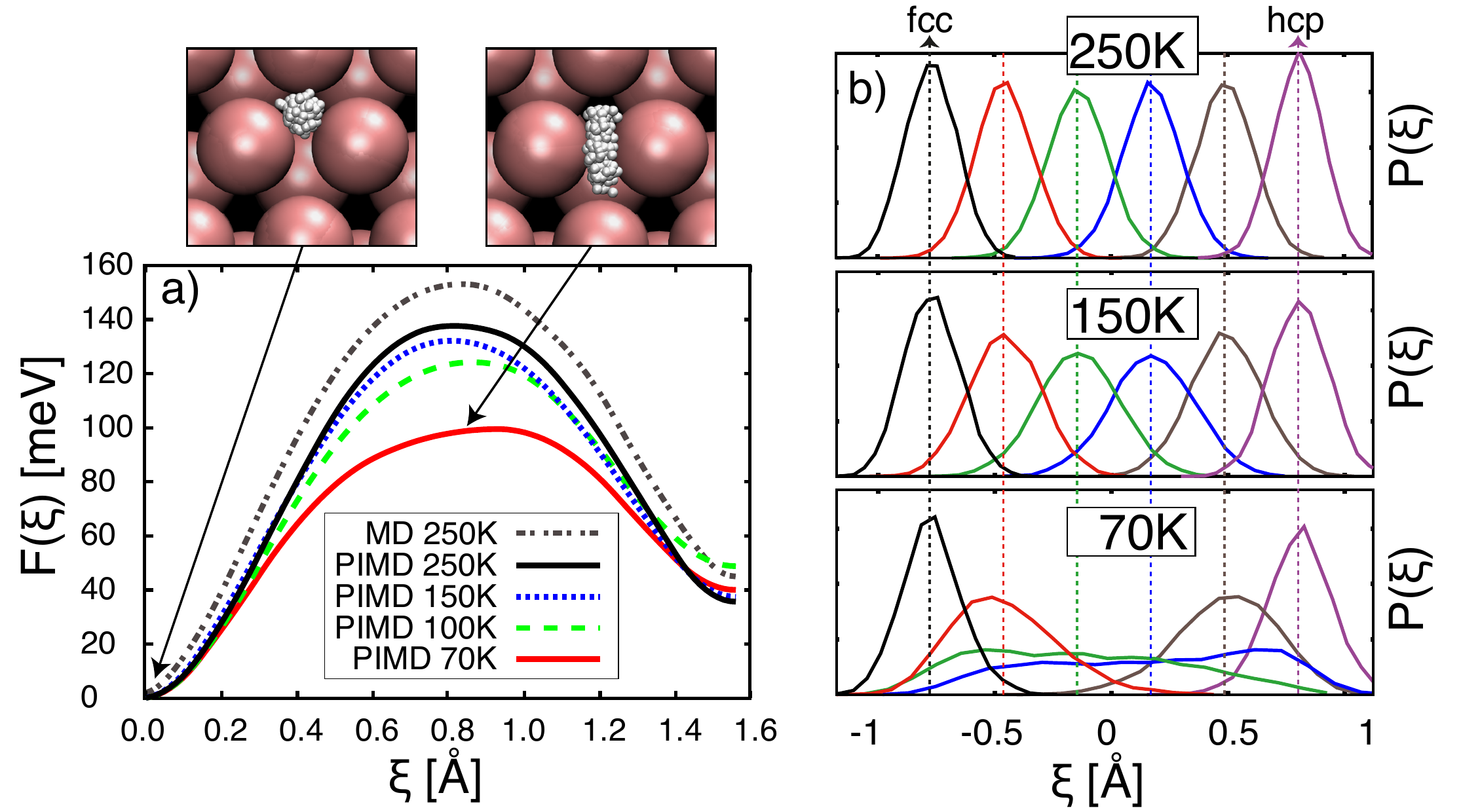}
 \caption{\label{fig:freenrj} a) Free energy barriers for H 
diffusion from fcc to hcp sites from ab initio PIMD compared to the temperature-independent barrier from classical-nucleus ab initio MD.
Above, several simulation snapshots are superimposed to show 
the distribution of beads when the centroid position is constrained 
at the initial site (fcc, left) and at the transition state (bridge, right) at 70~K.
b) Projected H atom bead probability distributions at a selected series of fixed centroid positions along the fcc-hcp path. Dashed vertical lines indicate the corresponding fixed centroid positions.}
 \end{figure}

\ref{fig:freenrj}a) shows quantum free energy barriers from PIMD at different temperatures and, for comparison, results of classical-nucleus MD simulations at 250~K. 
The free energy barrier obtained with classical nuclei is at $\sim$150 meV very similar to the underlying potential energy barrier. 
At the same temperature the free energy barrier obtained from PIMD is only slightly lower (by $\sim$15~meV) due mainly to ZPE effects.
That the slight lowering of the barrier predominantly comes from ZPE effects and not tunneling can be seen from panel b), where the width of the H probability distribution is plotted as a function of H position along the reaction coordinate. At  250~K the width of the the H distribution is unaffected by the (fixed) position of the H centroid along the fcc-hcp path.  However, as the temperature is lowered the H probability distribution broadens and the quantum free energy barrier drops. At 70~K the quantum free energy barrier is 99 meV, ~65\% of the original classical barrier.  This substantial reduction of the quantum free energy barrier arises from tunneling of the H through the potential barrier, as can be seen from the partially bimodal nature of the H probability distribution function (\ref{fig:freenrj}b)). 

\ref{fig:rate} reveals two main effects from the quantum treatment of nuclei in PIMD. Firstly, the crossover to tunneling-dominated diffusion is reflected by the changes in the PIMD rate, with a gradual change of slope in the Arrhenius plot below 100~K.  Secondly, at intermediate to high temperatures the gradient from PIMD agrees well with the experiment due to the decrease of the quantum free energy barrier and the inclusion of ZPE effects, highlighting the importance of quantum nuclear effects even at high temperatures. Indeed, fitting the same Arrhenius expression to the PIMD results as used for the experimental rates we obtain $E_a = 105$~meV, close to the experimental value of 95~meV.
However, PIMD gives a larger prefactor than the experiment resulting in rates faster by around a factor of 4 at high temperatures. This is likely to be due to the use of the classical velocity in the expression for the rate (see experimental methods).  To illustrate the agreement between the experimental and PIMD gradients at high temperatures we show in the inset of \ref{fig:rate} theoretical rates that have been scaled by 0.25. 

Despite the overall qualitative agreement with experiment, it is apparent that at low temperatures the PIMD results underestimate the experimental rate by 2--3 orders of magnitude.  
This could potentially be explained by the difference between the calculated and experimental estimates of the fcc-hcp energy difference and activation energy. 
To investigate this discrepancy in the low temperature rates we use a more flexible and computationally efficient approach than PIMD, specifically instanton theory~\cite{andersson2009comparison}.
For the instanton calculations we used a 1D potential represented by a polynomial fitted to the underlying DFT potential energy surface along the fcc-hcp diffusion path.
Initially we validated the instanton calculations by comparing the rates it produced with those from PIMD, and as shown in the low temperature region of \ref{fig:rate} the instanton and PIMD results agree quite well. 
By construction, instantons only provide rates in the tunneling regime.
Using the instanton approach we then investigated the sensitivity of the computed rates to the fcc-hcp energy difference. 
To this end we modified the 1D DFT potential by reducing the activation energy and hence also the fcc-hcp energy difference by 25~meV, bringing the binding site energy difference into close agreement with the experimental result. 
The dashed lines in \ref{fig:rate} show instanton rates for this modified potential. Indeed, 
significantly better agreement with the low-temperature experimental rates is now seen. The 
smaller gradient, due to the smaller fcc-hcp energy difference and thus smaller lattice deformation 
energy required for energy coincidence between fcc and hcp vibrational states, 
agrees better with the nearly temperature independent 
experimental rate, and the magnitude of tunneling is strongly enhanced by the small reduction in the
activation energy.
These results highlight the large sensitivity of computed jump rates on details of the PES, 
and hence the importance of performing accurate calculations. 
On the other hand, the rather good 
agreement between instanton calculations and PIMD rates on the unmodified PES show that polaron 
effects play a minor role on this surface, in agreement with the experimental observation of low
adsorbate-surface friction. 


In summary, we have presented a combined experimental and theoretical study of the diffusion of H on Ru(0001), showing a transition from overbarrier hopping between fcc and hcp adsorption sites at high temperatures to quantum tunneling at low temperatures.  The experimental results using HeSE give evidence for low adsorbate-substrate friction and repulsive inter-adsorbate interactions, and our experimental tunneling rates for H are much higher than those seen previously for other surfaces such as Ni(111)~\cite{lee1993observation,gomer1991fem,cao1997HNi111}, Cu(111)~\cite{sykes2012} and Cu(001)~\cite{ho2000stm}.
Tunneling from excited vibrational states was found to play a key role for H diffusion on Ni(111)~\cite{badescu2001quantum}. 
For H on Ru(0001), however, the low temperature rate in \ref{fig:rate} is effectively non-activated (the data is fitted by an activation energy of (-2$\pm$2)~meV~\cite{supmat} and so tunneling from the excited states (the lowest of which has been determined from HREELS to be 84~meV~\cite{kostov2004hreels}) does not contribute significantly at low temperatures.  On the other hand, the fcc-hcp energy barrier is 150~meV for Ru(0001)
but closer to 200~meV for Ni(111)~\cite{cao1997HNi111}, which would lead to faster tunneling for the case of H on Ru(0001) in the ground state.  
Recent STM results for H diffusing on Cu(111) at 5~K~\cite{sykes2012} (comprising streaks on STM images) were interpreted to give an H jump rate of 30~Hz.  However, in addition to the possible presence of tip induced effects, the imaging rate was only 5.5 lines~s$^{-1}$ and so would be insensitive to diffusion as fast as reported here.

Ab initio PIMD rates show reasonable agreement with the temperature dependence of the experimental results at high temperatures while the tunneling rates at low temperatures are significantly underestimated, similar to 
previous studies on other surfaces employing empirical potentials~\cite{mattsson1995qtstHNi,baer1998quantum,suleimanov2012HNiRPMD}. 
This discrepancy is highly interesting and 
by performing instanton calculations on a modified potential energy surface we  
showed that small changes in activation energy and energy difference between fcc and
hcp adsorption sites may play a role. However, other features of this system may 
also be important in driving the faster experimental than theoretical rates which 
highlights the need for future experiments and theoretical calculations of H diffusion on other well-defined metal surfaces. 

\section{Experimental}

Experimental measurements were carried out using the Cambridge HeSE Spectrometer~\cite{jardine2009helium}, which measures the Intermediate Scattering Function (ISF) $I\left(\Delta{\mathbf{K}},t\right)$, related to the dynamic structure factor $S\left(\Delta\mathbf{K},\Delta \omega\right)$ and the pair correlation function $G\left(\mathbf{R},t\right)$ by temporal and spatial Fourier transforms respectively~\cite{jardine2009helium}.
As such, HeSE is a uniquely non-invasive technique that can be used to study the detailed mechanism of individual atomic jumps on well-defined surfaces.
The single crystal Ru(0001) sample (Surface Prep. Lab., The Netherlands) was cleaned by repeated cycles of argon sputtering, flash annealing, oxidation, and reduction \cite{Marchini2007}, and the surface quality was monitored using helium reflectivity measurements.  
Molecular H (Air Liquide, 99.999$\%$ purity) was dosed by backfilling the scattering chamber, and surface uptake of atomic H was followed using the specularly scattered helium beam to achieve a coverage of 0.2~monolayer (ML), determined from the known dose of H and sticking probability \cite{feulnerandmenzel}.  

All DFT calculations were performed within a periodic plane-wave framework using projector-augmented wave potentials in the VASP code~\cite{kresse1996vasp1,kresse1999ultrasoft} modified for PIMD simulations~\cite{alfe2010ab}.  
The PIMD simulations were performed with 16 replicas (\textit{i.e.} beads of the ring-polymers).
Tests with 32 and 64 beads showed very good agreement with the 16 bead simulations (see Supporting Information for computational details~\cite{supmat}).

PIMD-based QTST~\cite{gillan1987quantum,voth1989qtst} gives the jump rate over the reaction coordinate $\xi$ (the fcc-hcp diffusion path) as
\begin{equation}
\label{eq:kqtst}
k_{QTST}(T) = \frac{\bar{v}}{2} P(\xi_{\ddagger},T) =\frac{\bar{v}}{2} P(\xi_{0},T)e^{-\Delta F/k_BT},
\end{equation}
where $\bar{v}$ is the thermal velocity: $\bar{v}=\sqrt{2k_{B}T/\pi m}$. $P(\xi_{\ddagger},T)$ and $P(\xi_{0},T)$ are the centroid (center of mass of the ring-polymer) probability densities at the transition state $\xi_{\ddagger}$ and initial (fcc) state $\xi_0$, respectively, and $\Delta F$ is the corresponding free energy difference obtained from a series of constrained-centroid PIMD simulations~\cite{supmat}. 
Although representing the reaction coordinate in terms of centroid positions becomes inaccurate for highly asymmetric potentials at temperatures far below the crossover to quantum tunneling~\cite{makarov1995quantum,mills1997generalized,richardson2009ring}, it should provide an accurate description for the moderately asymmetric potential just below the crossover temperature as studied here.

\acknowledgement

The authors would like to thank A.~P. Jardine, M.~I.~J. Probert, P. Hasnip and J. Aarons for helpful discussions.  E.~M.~M. is supported by an EPSRC studentship.  K.~T.~W. and A.~M. are supported by the European Research Council. A.~M. is also supported by the Royal Society through a Royal Society Wolfson Research Merit Award. This work made use of the Cambridge Spin-Echo Spectrometer and, through membership of the UK's HPC Materials Chemistry Consortium which is funded by the EPSRC (EP/F067496), the HECToR supercomputer. 

\suppinfo
Further experimental and computational details are provided in the Supporting Information.
This material is available free of charge via the Internet at http://pubs.ac.org.


\begin{mcitethebibliography}{48}
\providecommand*{\natexlab}[1]{#1}
\providecommand*{\mciteSetBstSublistMode}[1]{}
\providecommand*{\mciteSetBstMaxWidthForm}[2]{}
\providecommand*{\mciteBstWouldAddEndPuncttrue}
  {\def\EndOfBibitem{\unskip.}}
\providecommand*{\mciteBstWouldAddEndPunctfalse}
  {\let\EndOfBibitem\relax}
\providecommand*{\mciteSetBstMidEndSepPunct}[3]{}
\providecommand*{\mciteSetBstSublistLabelBeginEnd}[3]{}
\providecommand*{\EndOfBibitem}{}
\mciteSetBstSublistMode{f}
\mciteSetBstMaxWidthForm{subitem}{(\alph{mcitesubitemcount})}
\mciteSetBstSublistLabelBeginEnd{\mcitemaxwidthsubitemform\space}
{\relax}{\relax}

\bibitem[Li et~al.(2011)Li, Walker, and Michaelides]{li2011quantum}
Li,~X.~Z.; Walker,~B.; Michaelides,~A. \emph{Proc. Natl. Acad. Sci. (USA)}
  \textbf{2011}, \emph{108}, 6369\relax
\mciteBstWouldAddEndPuncttrue
\mciteSetBstMidEndSepPunct{\mcitedefaultmidpunct}
{\mcitedefaultendpunct}{\mcitedefaultseppunct}\relax
\EndOfBibitem
\bibitem[Garcia-Viloca et~al.(2004)Garcia-Viloca, Gao, Karplus, and
  Truhlar]{garcia2004enzymes}
Garcia-Viloca,~M.; Gao,~J.; Karplus,~M.; Truhlar,~D.~G. \emph{Science}
  \textbf{2004}, \emph{303}, 186\relax
\mciteBstWouldAddEndPuncttrue
\mciteSetBstMidEndSepPunct{\mcitedefaultmidpunct}
{\mcitedefaultendpunct}{\mcitedefaultseppunct}\relax
\EndOfBibitem
\bibitem[Koval et~al.(2002)Koval, Kohanoff, Migoni, and
  Tosatti]{koval2002ferro}
Koval,~S.; Kohanoff,~J.; Migoni,~R.~L.; Tosatti,~E. \emph{Phys. Rev. Lett.}
  \textbf{2002}, \emph{89}, 187602\relax
\mciteBstWouldAddEndPuncttrue
\mciteSetBstMidEndSepPunct{\mcitedefaultmidpunct}
{\mcitedefaultendpunct}{\mcitedefaultseppunct}\relax
\EndOfBibitem
\bibitem[Benoit et~al.(1998)Benoit, Marx, and Parrinello]{benoit1998tunnelling}
Benoit,~M.; Marx,~D.; Parrinello,~M. \emph{Nature} \textbf{1998}, \emph{392},
  258\relax
\mciteBstWouldAddEndPuncttrue
\mciteSetBstMidEndSepPunct{\mcitedefaultmidpunct}
{\mcitedefaultendpunct}{\mcitedefaultseppunct}\relax
\EndOfBibitem
\bibitem[Li et~al.(2010)Li, Probert, Alavi, and Michaelides]{li2010quantum}
Li,~X.~Z.; Probert,~M. I.~J.; Alavi,~A.; Michaelides,~A. \emph{Phys. Rev.
  Lett.} \textbf{2010}, \emph{104}, 066102\relax
\mciteBstWouldAddEndPuncttrue
\mciteSetBstMidEndSepPunct{\mcitedefaultmidpunct}
{\mcitedefaultendpunct}{\mcitedefaultseppunct}\relax
\EndOfBibitem
\bibitem[Fukai(2005)]{fukai2005metal}
Fukai,~Y. \emph{The metal-hydrogen system: basic bulk properties};
\newblock Springer Verlag, 2005;
\newblock Vol.~21\relax
\mciteBstWouldAddEndPuncttrue
\mciteSetBstMidEndSepPunct{\mcitedefaultmidpunct}
{\mcitedefaultendpunct}{\mcitedefaultseppunct}\relax
\EndOfBibitem
\bibitem[Ala-Nissila et~al.(2002)Ala-Nissila, Ferrando, and
  Ying]{alanissila2002surfdiffus}
Ala-Nissila,~T.; Ferrando,~R.; Ying,~S.~C. \emph{Adv. Phys.} \textbf{2002},
  \emph{51}, 949\relax
\mciteBstWouldAddEndPuncttrue
\mciteSetBstMidEndSepPunct{\mcitedefaultmidpunct}
{\mcitedefaultendpunct}{\mcitedefaultseppunct}\relax
\EndOfBibitem
\bibitem[Lin and Gomer(1991)]{gomer1991fem}
Lin,~T.~S.; Gomer,~R. \emph{Surf. Sci.} \textbf{1991}, \emph{225}, 41\relax
\mciteBstWouldAddEndPuncttrue
\mciteSetBstMidEndSepPunct{\mcitedefaultmidpunct}
{\mcitedefaultendpunct}{\mcitedefaultseppunct}\relax
\EndOfBibitem
\bibitem[Lee et~al.(1993)Lee, Zhu, Wong, Deng, and Linke]{lee1993observation}
Lee,~A.; Zhu,~X.~D.; Wong,~A.; Deng,~L.; Linke,~U. \emph{Phys. Rev. B}
  \textbf{1993}, \emph{48}, 11256\relax
\mciteBstWouldAddEndPuncttrue
\mciteSetBstMidEndSepPunct{\mcitedefaultmidpunct}
{\mcitedefaultendpunct}{\mcitedefaultseppunct}\relax
\EndOfBibitem
\bibitem[Cao et~al.(1997)Cao, Nabighian, and Zhu]{cao1997HNi111}
Cao,~G.~X.; Nabighian,~E.; Zhu,~X.~D. \emph{Phys. Rev. Lett.} \textbf{1997},
  \emph{79}, 3696\relax
\mciteBstWouldAddEndPuncttrue
\mciteSetBstMidEndSepPunct{\mcitedefaultmidpunct}
{\mcitedefaultendpunct}{\mcitedefaultseppunct}\relax
\EndOfBibitem
\bibitem[Badescu et~al.(2001)Badescu, Ying, and
  Ala-Nissila]{badescu2001quantum}
Badescu,~S.; Ying,~S.; Ala-Nissila,~T. \emph{Phys. Rev. Lett.} \textbf{2001},
  \emph{86}, 5092--5095\relax
\mciteBstWouldAddEndPuncttrue
\mciteSetBstMidEndSepPunct{\mcitedefaultmidpunct}
{\mcitedefaultendpunct}{\mcitedefaultseppunct}\relax
\EndOfBibitem
\bibitem[Lauhon and Ho(2000)]{ho2000stm}
Lauhon,~L.~J.; Ho,~W. \emph{Phys. Rev. Lett.} \textbf{2000}, \emph{85},
  4566\relax
\mciteBstWouldAddEndPuncttrue
\mciteSetBstMidEndSepPunct{\mcitedefaultmidpunct}
{\mcitedefaultendpunct}{\mcitedefaultseppunct}\relax
\EndOfBibitem
\bibitem[Kua et~al.(2001)Kua, Lauhon, Ho, and Goddard~III]{kua2001direct}
Kua,~J.; Lauhon,~L.~J.; Ho,~W.; Goddard~III,~W.~A. \emph{J. Chem. Phys.}
  \textbf{2001}, \emph{115}, 5620\relax
\mciteBstWouldAddEndPuncttrue
\mciteSetBstMidEndSepPunct{\mcitedefaultmidpunct}
{\mcitedefaultendpunct}{\mcitedefaultseppunct}\relax
\EndOfBibitem
\bibitem[Sundell and Wahnstr\"om(2004)]{sundell2004HDCu001}
Sundell,~P.~G.; Wahnstr\"om,~G. \emph{Phys. Rev. B} \textbf{2004}, \emph{70},
  081403\relax
\mciteBstWouldAddEndPuncttrue
\mciteSetBstMidEndSepPunct{\mcitedefaultmidpunct}
{\mcitedefaultendpunct}{\mcitedefaultseppunct}\relax
\EndOfBibitem
\bibitem[Sundell and Wahnstr\"om(2005)]{sundell2005HDCu001}
Sundell,~P.~G.; Wahnstr\"om,~G. \emph{Surf. Sci} \textbf{2005}, \emph{593},
  102\relax
\mciteBstWouldAddEndPuncttrue
\mciteSetBstMidEndSepPunct{\mcitedefaultmidpunct}
{\mcitedefaultendpunct}{\mcitedefaultseppunct}\relax
\EndOfBibitem
\bibitem[Feulner and Menzel(1985)]{feulnerandmenzel}
Feulner,~P.; Menzel,~D. \emph{Surf. Sci.} \textbf{1985}, \emph{154},
  465--488\relax
\mciteBstWouldAddEndPuncttrue
\mciteSetBstMidEndSepPunct{\mcitedefaultmidpunct}
{\mcitedefaultendpunct}{\mcitedefaultseppunct}\relax
\EndOfBibitem
\bibitem[Mak et~al.(1986)Mak, Brand, Deckert, and George]{mak1986LITD}
Mak,~C.; Brand,~J.; Deckert,~A.; George,~S. \emph{J. Chem. Phys.}
  \textbf{1986}, \emph{85}, 1676--1680\relax
\mciteBstWouldAddEndPuncttrue
\mciteSetBstMidEndSepPunct{\mcitedefaultmidpunct}
{\mcitedefaultendpunct}{\mcitedefaultseppunct}\relax
\EndOfBibitem
\bibitem[Lindroos et~al.(1987)Lindroos, Pfn{\"u}r, Feulner, and
  Menzel]{lindroos1987study}
Lindroos,~M.; Pfn{\"u}r,~H.; Feulner,~P.; Menzel,~D. \emph{Surf. Sci.}
  \textbf{1987}, \emph{180}, 237--251\relax
\mciteBstWouldAddEndPuncttrue
\mciteSetBstMidEndSepPunct{\mcitedefaultmidpunct}
{\mcitedefaultendpunct}{\mcitedefaultseppunct}\relax
\EndOfBibitem
\bibitem[Braun et~al.(1997)Braun, Kostov, Witte, Surnev, Skofronick, Safron,
  and W\"oll]{Braun1992}
Braun,~J.; Kostov,~K.~L.; Witte,~G.; Surnev,~L.; Skofronick,~J.~G.;
  Safron,~S.~A.; W\"oll,~C. \emph{Surf. Sci.} \textbf{1997}, \emph{372},
  132--144\relax
\mciteBstWouldAddEndPuncttrue
\mciteSetBstMidEndSepPunct{\mcitedefaultmidpunct}
{\mcitedefaultendpunct}{\mcitedefaultseppunct}\relax
\EndOfBibitem
\bibitem[Sandhoff et~al.(1993)Sandhoff, Pfn\"ur, and Everts]{sandhoff1993MC}
Sandhoff,~M.; Pfn\"ur,~H.; Everts,~H. \emph{Surf. Sci.} \textbf{1993},
  \emph{280}, 185\relax
\mciteBstWouldAddEndPuncttrue
\mciteSetBstMidEndSepPunct{\mcitedefaultmidpunct}
{\mcitedefaultendpunct}{\mcitedefaultseppunct}\relax
\EndOfBibitem
\bibitem[Shi and Jacobi(1994)]{Shi1994vibrations}
Shi,~H.; Jacobi,~K. \emph{Surf. Sci.} \textbf{1994}, \emph{313}, 289--294\relax
\mciteBstWouldAddEndPuncttrue
\mciteSetBstMidEndSepPunct{\mcitedefaultmidpunct}
{\mcitedefaultendpunct}{\mcitedefaultseppunct}\relax
\EndOfBibitem
\bibitem[Kostov et~al.(2004)Kostov, Widdra, and Menzel]{kostov2004hreels}
Kostov,~K.~L.; Widdra,~W.; Menzel,~D. \emph{Surf. Sci.} \textbf{2004},
  \emph{560}, 130\relax
\mciteBstWouldAddEndPuncttrue
\mciteSetBstMidEndSepPunct{\mcitedefaultmidpunct}
{\mcitedefaultendpunct}{\mcitedefaultseppunct}\relax
\EndOfBibitem
\bibitem[Tatarkhanov et~al.(2008)Tatarkhanov, Rose, Fomin, Ogletree, and
  Salmeron]{tatarkhanov2008hydrogen}
Tatarkhanov,~M.; Rose,~F.; Fomin,~E.; Ogletree,~D.~F.; Salmeron,~M. \emph{Surf.
  Sci.} \textbf{2008}, \emph{602}, 487--492\relax
\mciteBstWouldAddEndPuncttrue
\mciteSetBstMidEndSepPunct{\mcitedefaultmidpunct}
{\mcitedefaultendpunct}{\mcitedefaultseppunct}\relax
\EndOfBibitem
\bibitem[Lizzit et~al.(2009)Lizzit, Zhang, Kostov, Petaccia, Baraldi, Menzel,
  and Reuter]{lizzit2009Ru0001core}
Lizzit,~S.; Zhang,~Y.; Kostov,~K.~L.; Petaccia,~L.; Baraldi,~A.; Menzel,~D.;
  Reuter,~K. \emph{J Phys.: Cond. Matter} \textbf{2009}, \emph{21},
  134009\relax
\mciteBstWouldAddEndPuncttrue
\mciteSetBstMidEndSepPunct{\mcitedefaultmidpunct}
{\mcitedefaultendpunct}{\mcitedefaultseppunct}\relax
\EndOfBibitem
\bibitem[Chou and Chelikowsky(1987)]{chelikowsky1987HRu}
Chou,~M.~Y.; Chelikowsky,~J.~R. \emph{Phys. Rev. Lett.} \textbf{1987},
  \emph{59}, 1737\relax
\mciteBstWouldAddEndPuncttrue
\mciteSetBstMidEndSepPunct{\mcitedefaultmidpunct}
{\mcitedefaultendpunct}{\mcitedefaultseppunct}\relax
\EndOfBibitem
\bibitem[Xu et~al.(2005)Xu, Xiao, and Zu]{xu2005HRu0001castep}
Xu,~L.; Xiao,~H.~Y.; Zu,~X.~T. \emph{Chem. Phys.} \textbf{2005}, \emph{315},
  155\relax
\mciteBstWouldAddEndPuncttrue
\mciteSetBstMidEndSepPunct{\mcitedefaultmidpunct}
{\mcitedefaultendpunct}{\mcitedefaultseppunct}\relax
\EndOfBibitem
\bibitem[Kristinsd\'ottir and Sk\'ulason(2006)]{KristinsdottirDFT}
Kristinsd\'ottir,~L.; Sk\'ulason,~E. \emph{Surf. Sci.} \textbf{2012},
  \emph{606}, 1400--1404\relax
\mciteBstWouldAddEndPuncttrue
\mciteSetBstMidEndSepPunct{\mcitedefaultmidpunct}
{\mcitedefaultendpunct}{\mcitedefaultseppunct}\relax
\EndOfBibitem
\bibitem[Jardine et~al.(2009)Jardine, Hedgeland, Alexandrowicz, Allison, and
  Ellis]{jardine2009helium}
Jardine,~A.~P.; Hedgeland,~H.; Alexandrowicz,~G.; Allison,~W.; Ellis,~J.
  \emph{Prog. Surf. Sci.} \textbf{2009}, \emph{84}, 323 -- 379\relax
\mciteBstWouldAddEndPuncttrue
\mciteSetBstMidEndSepPunct{\mcitedefaultmidpunct}
{\mcitedefaultendpunct}{\mcitedefaultseppunct}\relax
\EndOfBibitem
\bibitem[Gillan(1987)]{gillan1987quantum}
Gillan,~M.~J. \emph{J. Phys. C: Solid State Phys.} \textbf{1987}, \emph{20},
  3621\relax
\mciteBstWouldAddEndPuncttrue
\mciteSetBstMidEndSepPunct{\mcitedefaultmidpunct}
{\mcitedefaultendpunct}{\mcitedefaultseppunct}\relax
\EndOfBibitem
\bibitem[Voth et~al.(1989)Voth, Chandler, and Miller]{voth1989qtst}
Voth,~G.~A.; Chandler,~D.; Miller,~W.~H. \emph{J. Chem. Phys.} \textbf{1989},
  \emph{91}, 7749\relax
\mciteBstWouldAddEndPuncttrue
\mciteSetBstMidEndSepPunct{\mcitedefaultmidpunct}
{\mcitedefaultendpunct}{\mcitedefaultseppunct}\relax
\EndOfBibitem
\bibitem[Jewell et~al.(2012)Jewell, Peng, Mattera, Lewis, Murphy, Kyriakou,
  Mavrikakis, and Sykes]{sykes2012}
Jewell,~A.~D.; Peng,~G.; Mattera,~M. F.~G.; Lewis,~E.~A.; Murphy,~C.~J.;
  Kyriakou,~G.; Mavrikakis,~M.; Sykes,~E. C.~H. \emph{ACS Nano} \textbf{2012},
  \emph{0}, null\relax
\mciteBstWouldAddEndPuncttrue
\mciteSetBstMidEndSepPunct{\mcitedefaultmidpunct}
{\mcitedefaultendpunct}{\mcitedefaultseppunct}\relax
\EndOfBibitem
\bibitem[Jardine et~al.(2010)Jardine, Lee, Ward, Alexandrowicz, Hedgeland,
  Allison, Ellis, and Pollak]{jardine2010HPt111_HeSE}
Jardine,~A.~P.; Lee,~E. Y.~M.; Ward,~D.~J.; Alexandrowicz,~G.; Hedgeland,~H.;
  Allison,~W.; Ellis,~J.; Pollak,~E. \emph{Phys. Rev. Lett.} \textbf{2010},
  \emph{105}, 136101\relax
\mciteBstWouldAddEndPuncttrue
\mciteSetBstMidEndSepPunct{\mcitedefaultmidpunct}
{\mcitedefaultendpunct}{\mcitedefaultseppunct}\relax
\EndOfBibitem
\bibitem[Andersson et~al.(2009)Andersson, Nyman, Arnaldsson, Manthe, and
  J{\'o}nsson]{andersson2009comparison}
Andersson,~S.; Nyman,~G.; Arnaldsson,~A.; Manthe,~U.; J{\'o}nsson,~H. \emph{J.
  Phys. Chem. A} \textbf{2009}, \emph{113}, 4468--4478\relax
\mciteBstWouldAddEndPuncttrue
\mciteSetBstMidEndSepPunct{\mcitedefaultmidpunct}
{\mcitedefaultendpunct}{\mcitedefaultseppunct}\relax
\EndOfBibitem
\bibitem[Tuddenham et~al.(2010)Tuddenham, Hedgeland, Jardine, Lechner, Hinch,
  and Allison]{tuddenham2010lineshapes}
Tuddenham,~F.; Hedgeland,~H.; Jardine,~A.; Lechner,~B.; Hinch,~B.; Allison,~W.
  \emph{Surf. Sci.} \textbf{2010}, \emph{604}, 1459--1475\relax
\mciteBstWouldAddEndPuncttrue
\mciteSetBstMidEndSepPunct{\mcitedefaultmidpunct}
{\mcitedefaultendpunct}{\mcitedefaultseppunct}\relax
\EndOfBibitem
\bibitem[{S}ee {S}upporting{I}nformation()]{supmat}
{S}ee {S}upporting{I}nformation,\relax
\mciteBstWouldAddEndPuncttrue
\mciteSetBstMidEndSepPunct{\mcitedefaultmidpunct}
{\mcitedefaultendpunct}{\mcitedefaultseppunct}\relax
\EndOfBibitem
\bibitem[Jardine et~al.(2007)Jardine, Alexandrowicz, Hedgeland, Diehl, Allison,
  and Ellis]{jardineCs}
Jardine,~A.~P.; Alexandrowicz,~G.; Hedgeland,~H.; Diehl,~R.~D.; Allison,~W.;
  Ellis,~J. \emph{J. Phys. Cond. Matter} \textbf{2007}, \emph{19}, 305010\relax
\mciteBstWouldAddEndPuncttrue
\mciteSetBstMidEndSepPunct{\mcitedefaultmidpunct}
{\mcitedefaultendpunct}{\mcitedefaultseppunct}\relax
\EndOfBibitem
\bibitem[Alexandrowicz et~al.(2006)Alexandrowicz, Jardine, Hedgeland, Allison,
  and Ellis]{GilNaCu}
Alexandrowicz,~G.; Jardine,~A.~P.; Hedgeland,~H.; Allison,~W.; Ellis,~J.
  \emph{Phys. Rev. Lett.} \textbf{2006}, \emph{97}, 156103\relax
\mciteBstWouldAddEndPuncttrue
\mciteSetBstMidEndSepPunct{\mcitedefaultmidpunct}
{\mcitedefaultendpunct}{\mcitedefaultseppunct}\relax
\EndOfBibitem
\bibitem[Mattsson and Wahnstr\"{o}m(1995)]{mattsson1995qtstHNi}
Mattsson,~T.~R.; Wahnstr\"{o}m,~G. \emph{Phys. Rev. B} \textbf{1995},
  \emph{51}, 1885\relax
\mciteBstWouldAddEndPuncttrue
\mciteSetBstMidEndSepPunct{\mcitedefaultmidpunct}
{\mcitedefaultendpunct}{\mcitedefaultseppunct}\relax
\EndOfBibitem
\bibitem[Baer et~al.(1998)Baer, Zeiri, and Kosloff]{baer1998quantum}
Baer,~R.; Zeiri,~Y.; Kosloff,~R. \emph{Surf. Sci.} \textbf{1998}, \emph{411},
  L783--L788\relax
\mciteBstWouldAddEndPuncttrue
\mciteSetBstMidEndSepPunct{\mcitedefaultmidpunct}
{\mcitedefaultendpunct}{\mcitedefaultseppunct}\relax
\EndOfBibitem
\bibitem[Suleimanov(2012)]{suleimanov2012HNiRPMD}
Suleimanov,~Y.~V. \emph{J. Phys. Chem. C} \textbf{2012}, \emph{116},
  11141\relax
\mciteBstWouldAddEndPuncttrue
\mciteSetBstMidEndSepPunct{\mcitedefaultmidpunct}
{\mcitedefaultendpunct}{\mcitedefaultseppunct}\relax
\EndOfBibitem
\bibitem[Marchini et~al.(2007)Marchini, Gunther, and Wintterlin]{Marchini2007}
Marchini,~S.; Gunther,~S.; Wintterlin,~J. \emph{Phys. Rev. B} \textbf{2007},
  \emph{76}, 075429\relax
\mciteBstWouldAddEndPuncttrue
\mciteSetBstMidEndSepPunct{\mcitedefaultmidpunct}
{\mcitedefaultendpunct}{\mcitedefaultseppunct}\relax
\EndOfBibitem
\bibitem[Kresse and Furthm\"uller(1996)]{kresse1996vasp1}
Kresse,~G.; Furthm\"uller,~J. \emph{J. Comp. Mater. Sci.} \textbf{1996},
  \emph{6}, 15--50\relax
\mciteBstWouldAddEndPuncttrue
\mciteSetBstMidEndSepPunct{\mcitedefaultmidpunct}
{\mcitedefaultendpunct}{\mcitedefaultseppunct}\relax
\EndOfBibitem
\bibitem[Kresse and Joubert(1999)]{kresse1999ultrasoft}
Kresse,~G.; Joubert,~D. \emph{Phys. Rev. B} \textbf{1999}, \emph{59},
  1758\relax
\mciteBstWouldAddEndPuncttrue
\mciteSetBstMidEndSepPunct{\mcitedefaultmidpunct}
{\mcitedefaultendpunct}{\mcitedefaultseppunct}\relax
\EndOfBibitem
\bibitem[Alf{\`e} and Gillan(2010)]{alfe2010ab}
Alf{\`e},~D.; Gillan,~M.~J. \emph{J. Chem. Phys.} \textbf{2010}, \emph{133},
  044103\relax
\mciteBstWouldAddEndPuncttrue
\mciteSetBstMidEndSepPunct{\mcitedefaultmidpunct}
{\mcitedefaultendpunct}{\mcitedefaultseppunct}\relax
\EndOfBibitem
\bibitem[Makarov and Topaler(1995)]{makarov1995quantum}
Makarov,~D.~E.; Topaler,~M. \emph{Phys. Rev. E} \textbf{1995}, \emph{52},
  178\relax
\mciteBstWouldAddEndPuncttrue
\mciteSetBstMidEndSepPunct{\mcitedefaultmidpunct}
{\mcitedefaultendpunct}{\mcitedefaultseppunct}\relax
\EndOfBibitem
\bibitem[Mills et~al.(1997)Mills, Schenter, Makarov, and
  J{\'o}nsson]{mills1997generalized}
Mills,~G.; Schenter,~G.~K.; Makarov,~D.~E.; J{\'o}nsson,~H. \emph{Chem. Phys.
  Lett.} \textbf{1997}, \emph{278}, 91\relax
\mciteBstWouldAddEndPuncttrue
\mciteSetBstMidEndSepPunct{\mcitedefaultmidpunct}
{\mcitedefaultendpunct}{\mcitedefaultseppunct}\relax
\EndOfBibitem
\bibitem[Richardson and Althorpe(2009)]{richardson2009ring}
Richardson,~J.; Althorpe,~S. \emph{J. Chem. Phys.} \textbf{2009}, \emph{131},
  214106\relax
\mciteBstWouldAddEndPuncttrue
\mciteSetBstMidEndSepPunct{\mcitedefaultmidpunct}
{\mcitedefaultendpunct}{\mcitedefaultseppunct}\relax
\EndOfBibitem
\end{mcitethebibliography}
\providecommand*{\mcitethebibliography}{\thebibliography}
\csname @ifundefined\endcsname{endmcitethebibliography}
{\let\endmcitethebibliography\endthebibliography}{}

\end{document}